\documentclass[sn-mathphys,Numbered,onecolumn]{sn-jnl}

\usepackage{graphicx}%
\usepackage{multirow}%
\usepackage{amsmath,amssymb,amsfonts}%
\usepackage{amsthm}%
\usepackage{float}
\usepackage{mathrsfs}%
\usepackage{comment}
\usepackage[title]{appendix}%
\usepackage{xcolor}%
\usepackage{textcomp}%
\usepackage{manyfoot}%
\usepackage{booktabs}%
\usepackage{algorithm}%
\usepackage{subcaption} 
\usepackage{algorithmicx}%
\usepackage{listings}%
\usepackage[noend]{algpseudocode}
\theoremstyle{thmstyleone}%
\theoremstyle{thmstylethree}%
\begin{document}


\title{Challenges and Solutions in Selecting Optimal  Lossless Data Compression Algorithms}
\author[1]{\fnm{Md. Atiqur } \sur{Rahman}}\email{atick.rasel@gmail.com}

\author[2]{\fnm{MM Fazle } \sur{Rabbi}}\email{rabbi@bubt.edu.bd}
\affil[1]{\orgdiv{\centering Department of Computer Science and Engineering}\\ \orgname{\centering East West University}, \orgaddress{\city{Dhaka}, \country{Bangladesh}}}
\affil[2]{\orgdiv{\centering Department of Computer Science and Engineering}\\ \orgname{\centering Bangladesh University of Business and Technology (BUBT)}, \orgaddress{\city{Dhaka}, \country{Bangladesh}}}

\abstract{
    The rapid growth of digital data has heightened the demand for efficient lossless compression methods. However, existing algorithms exhibit trade-offs: some achieve high compression ratios, others excel in encoding or decoding speed, and none consistently perform best across all dimensions. This mismatch complicates algorithm selection for applications where multiple performance metrics are simultaneously critical, such as medical imaging, which requires both compact storage and fast retrieval. To address this challenge, we present a mathematical framework that integrates compression ratio, encoding time, and decoding time into a unified performance score. The model normalizes and balances these metrics through a principled weighting scheme, enabling objective and fair comparisons among diverse algorithms. Extensive experiments on image and text datasets validate the approach, showing that it reliably identifies the most suitable compressor for different priority settings. Results also reveal that while modern learning-based codecs often provide superior compression ratios, classical algorithms remain advantageous when speed is paramount. The proposed framework offers a robust and adaptable decision-support tool for selecting optimal lossless data compression techniques, bridging theoretical measures with practical application needs.

}
\keywords{Lossless Data Compression, Algorithm Selection, Multi-criteria Optimization, Compression Ratio, Encoding Time, Decoding Time, Performance Evaluation.}

\maketitle

\section{Introduction}\label{sec1}
Over the past few decades, the rate at which data are generated has increased dramatically, resulting in significant processing costs \cite{bondvzulic2024simple, zhang2024learned,rahman2018histogram, masmoudi2014efficient,rahman2020burrows,li2024integrated,rahman2022integer}. Therefore, the need for data compression is greater than ever. Although lossy compression is commonly used in multimedia communication, achieving a higher compression ratio at the cost of quality degradation, certain data, such as artistic, scientific, and medical data, must be preserved intact \cite{rahman2024novel}. Thus, lossless data compression is important in many fields \cite{rhee2020channel}. Lossless data compression methods can be categorized as either learning-based or non-learning-based codecs. State-of-the-art non-learning-based techniques include Portable Network Graphics (PNG) \cite{fu2024hybrid,bhattacharjee2024efficient, boutell1997png,dai2021deep,qin2024urbanevolver,lin2021eapt}, Better Portable Graphics (BPG) \cite{bellard2015bpg,zhang2024practical,rahman2021text}, JPEG XR \cite{dufaux2009jpeg,rahman2023prediction}, JPEG 2000 \cite{rahman2018novel, rabbani2002jpeg2000,elmeligy2024evaluating,rahman2021pcbms}, JPEG-LS \cite{weinberger2000loco,rahman2019semi}, WebP \cite{weinberger2000loco}, LCIC \cite{version20001}, JPEG-XL \cite{azami2024lico,alakuijala2019jpeg}, Context-Based Adaptive Lossless Image Codec (CALIC) \cite{kim2013hierarchical}, AV1 Image File Format (AVIF) \cite{han2021technical}, H.264 \cite{wu1997context}, and FLIF \cite{sneyers2016flif, gumucs2024learned}. These techniques compress images by reducing spatial redundancy. Learning-based techniques include CBPNN \cite{schiopu2018macro}, MP-CNN \cite{schiopu2018residual}, REP-CNN \cite{mentzer2019practical}, L3C \cite{reed2017parallel}, MS-PixelCNN \cite{rhee2020channel}, CWPLIC \cite{rhee2020channel}, IDF \cite{hoogeboom2019integer}, PixelRNN \cite{van2016pixel}, and PixelCNN \cite{oord2016conditional,zhu2024clustering,chen2023mngnas}. Learning-based methods provide the technology needed to develop productive solutions with improved performance compared with classical techniques \cite{schiopu2020study}.

PixelCNN \cite{van2016pixel}, a deep convolutional network model, has been compared with deep diffusion [18], deep Gaussian mixture models \cite{han2021technical}, RIDE \cite{wiegand2003overview}, Row LSTM, and Diagonal BiLSTM on the CIFAR-10 dataset. Diagonal BiLSTM provided a higher compression ratio, whereas PixelCNN was the fastest architecture. MS-PixelCNN is 105 times faster than PixelCNN \cite{mentzer2019practical}. Mentzer et al. \cite{mentzer2019practical}  proposed a learning-based lossless image compression technique, L3C, and compared it with two learning methods (PixelCNN and MS-PixelCNN) and four classical techniques (PNG, JPEG 2000, WebP, and FLIF) on the ImageNet32 dataset \cite{krasin2017openimages}. L3C was found to be approximately $5.06 \times 10^2$ times faster than MS-PixelCNN and provided 20.6\% and 24.4\% more compression power than MS-PixelCNN and PixelCNN, respectively. The authors also compared L3C with PNG, JPEG 2000, WebP, and FLIF on $512 \times 512$ crops from the DIV2K dataset \cite{agustsson2017ntire}. FLIF achieved higher compression ratios than the other methods. L3C was approximately 8.08 times faster for encoding and 2.81 times slower for decoding than FLIF. However, JPEG 2000 and PNG provided faster encoding and decoding, respectively.

Rhee et al. \cite{rhee2020channel} proposed a context-adaptive lossless image compression technique using a multilayer perceptron (MLP) that requires less encoding and decoding time than a convolutional neural network (CNN). The authors compared it with several classical methods (PNG, JPEG 2000, WebP, LCIC, and FLIF) and a learning-based technique (L3C) on three high-resolution datasets: McMaster \cite{li2018learning}, Open Images \cite{krasin2017openimages}, and DIV2K \cite{agustsson2017ntire}. Compared with the classical codecs, the new method outperformed PNG, JPEG 2000, WebP, and LCIC across all datasets, whereas FLIF achieved compression to fewer bits per pixel (bpp) in the Open Images dataset. The new method also achieved a lower bpp than L3C in all datasets. However, it had the lowest encoding and decoding speeds among the methods, except that it offered a 13.71\% improvement over the encoding speed of L3C. Although methods based on CNNs and recurrent neural networks (RNNs) \cite{schiopu2018macro, reed2017parallel,rahman2022comparative, oord2016conditional, rippel2017real, schiopu2019cnn, rahman2018histogram,rahman2021lossless, dinh2014nice,germain2015made,dai2024deep,qian2024drac} achieved higher compression ratios than Rhee et al.'s method, they were impractical due to the extended time required for encoding and decoding: over an hour on a GPU. JPEG 2000 was the fastest in encoding, with their method being 99.88\% slower. Additionally, PNG and JPEG 2000 were both about 961 times faster than their method. RasterLite2 \cite{TransactionLink1} was used to apply various lossless compression techniques (LZ4 \cite{TransactionLink2}, DEFLATE \cite{deutsch1996rfc1951}, ZSTD \cite{TransactionLink3}, LZMA \cite{TransactionLink4}, PNG, WebP, and JPEG 2000) to the Orthophoto imagery (grayscale) dataset. Among these, WebP provided the highest compression ratio, LZ4 the fastest encoding, and PNG the fastest decoding.

In \cite{rhee2020channel,graves2013generating}, compression methods were compared on the basis of compression ratio, encoding time, or decoding time. However, in most other works \cite{masmoudi2014efficient,boutell1997png,dufaux2009jpeg,bellard2015bpg,dufaux2009jpeg,de2009subjective,rabbani2002jpeg2000,weinberger2000loco,version20001,kim2013hierarchical,alakuijala2019jpeg,sneyers2016flif,wu1997context,schiopu2019deep,schiopu2018macro,schiopu2018residual,mentzer2019practical,reed2017parallel,hoogeboom2019integer,van2016pixel,oord2016conditional,schiopu2020study,han2021technical,wiegand2003overview,zhang2011color,rippel2017real,schiopu2019cnn,dinh2014nice,germain2015made,chrabaszcz2017downsampled,collet2018zstandard,TransactionLink1,deutsch1996rfc1951,TransactionLink3,TransactionLink4,lin2020reversible,suzuki2018lossless,patwa2020semantic,alzahir2014innovative,crandall2014lossless,lee2017lossless,shen2016predictive,liu2018adaptive,karimi2012lossless}, only compression ratios have been used to evaluate compression techniques. Although some applications prioritize higher compression ratios, others prioritize faster encoding or decoding speeds. In many applications, however, two or all of these factors are important. This work thoroughly compares the performance of various compression techniques on publicly accessible datasets. Our research aims to provide a method to answer the following question for a given application: Which algorithm is the most efficient when considering compression ratio and encoding time, compression ratio and decoding time, encoding time and decoding time, or all three parameters equally?

This article is structured as follows: The purpose of the research is described in Section~\ref{Purposeofthisresearch}. The proposed methodology is introduced in Section~\ref{Proposedmethod}. In Section~\ref{ResultsandAnalysis}, experimental findings are presented and analyzed. In Section~\ref{Validatetheproposedmethod}, the effectiveness of the approach is demonstrated. Lastly, in Section~\ref{conclusions1}, we draw conclusions based on our study.

\section{Purpose} \label{Purposeofthisresearch}
The development of learning-based lossless data compression techniques is progressing rapidly. Such methods include L3C \cite{rhee2020channel}, deep Gaussian mixture models \cite{van2015locally,van2014factoring}, super-resolution-based compression (SReC) \cite{cao2020lossless}, CBPNN \cite{schiopu2019deep}, PixelRNN \cite{van2016pixel}, CWPLIC \cite{rhee2020channel}, deep diffusion \cite{sohl2015deep}, PixelCNN \cite{van2016conditional}, Gated PixelCNN \cite{van2016conditional}, REP-CNN \cite{schiopu2018residual}, MS-PixelCNN \cite{reed2017parallel}, IDF \cite{hoogeboom2019integer}, Row LSTM \cite{van2016pixel}, Diagonal BiLSTM \cite{van2016pixel}, LLICTI \cite{kamisli2022learned}, and MSPSM \cite{zhang2020lossless}. Additionally, many non-learning methods are widely used in various applications, such as the High-Efficiency Image File Format (HEIF) \cite{heikkila2016high}, lossless RLE compression of Amiga IFF images (ILBM) \cite{rusyn2016lossless}, the context-based, adaptive, lossless image codec (CALIC) \cite{wu1996calic}, lossless compression of black and white images (JBIG2) \cite{ye2001dictionary}, JPEG-LS \cite{weinberger2000loco}, JPEG XL \cite{alakuijala2020benchmarking}, JPEG XR \cite{dufaux2009jpeg}, TIFF \cite{sindhu2009images}, arithmetic coding \cite{rissanen1979arithmetic}, byte pair encoding \cite{shibata1999byte}, DEFLATE \cite{oswal2016deflate}, lZ4 \cite{bartik2015lz4}, LZMA \cite{parekar2014lossless}, LZO \cite{oberhumer2008lzo}, LZRW \cite{bartik2015lz4}, LZX \cite{salomon2004data}, and LCIC \cite{kim2013hierarchical}.

When working with heterogeneous datasets, a variety of learning and non-learning lossless compression techniques can be applied. These techniques are compared based on the compression ratio, encoding time, and decoding time. With advancements in technology, users' demands are increasing. Some applications require a lossless compression technique that provides a higher compression ratio or faster encoding or decoding processes. Selecting an algorithm solely on the basis of compression ratio, encoding time, or decoding time is straightforward. However, some applications may require a high compression ratio and a short encoding time, whereas others may need short encoding and decoding times, a high compression ratio with a short decoding time, or a combination of all three. Selecting the best algorithm in these cases is challenging. For example, for the DIV2K dataset \cite{mentzer2019practical}, Table~\ref{tab:my-table1} compares the average encoding and decoding times, as well as the bits per subpixel (bpsp), obtained in \cite{mentzer2019practical} for five different compressors: JPEG 2000, PNG, WebP, FLIF, and L3C. According to the authors, L3C offers a lower bpsp than JPEG 2000, PNG, and WebP. However, L3C's encoding process is much slower than these methods. Among the algorithms, FLIF offers the most compression power, but its encoding process is 85.93\%, 87.62\%, 99.14\%, and 90.87\% slower than L3C, PNG, JPEG 2000, and WebP, respectively. It can be seen from the table that JPEG 2000 and PNG achieve shorter encoding and decoding times, respectively. Therefore, it is difficult to identify the superior algorithm when two or all three performance measures are important. This research article proposes a mathematical method for choosing a superior lossless data compression algorithm on the basis of any possible combination of compression ratio, encoding time, and decoding time.

\begin{table*}[!h]
\centering
\caption{Experimental results of five different compressors on the DIV2K dataset, published in \cite{mentzer2019practical}}
\label{tab:my-table1}
\begin{tabular}{|l|c|c|c|}
\hline
\textbf{Compressor} &
  \textbf{\begin{tabular}[c]{@{}c@{}}Encoding  Time (s)\end{tabular}} &
  \textbf{\begin{tabular}[c]{@{}c@{}}Decoding  Time (s)\end{tabular}} &
  \textbf{bpsp} \\ \hline
JPFG 2000 & $1.48 \times 10^{-2}$ & $2.26 \times 10^{-4}$ & 3.471 \\ \hline
PNG       & 0.213                 & $6.09 \times 10^{-5}$ & 4.733 \\ \hline
WebP      & 0.157                 & $7.12 \times 10^{-2}$ & 3.447 \\ \hline
FLIF      & 1.72                  & 0.133                 & 3.291 \\ \hline
L3C       & 0.242                 & 0.374                 & 3.386 \\ \hline
\end{tabular}
\end{table*}

\section{Proposed Method} \label{Proposedmethod}
A fundamental principle of data compression is that the most efficient data compression algorithm is the one with the highest compression ratio $r$ combined with the shortest encoding time $e$ and decoding time $d$. This statement can be formulated  by expressing the performance $p$ of an algorithm as
\begin{equation}\label{equ:eq1_3}
p \propto 
\begin{cases}
r & \text{for } x = r, \\
\frac{1}{x} & \text{for } x \in \{e,d\}.
\end{cases}
\end{equation}
Let us assume that $n$ images are compressed using $m$ distinct lossless data compression techniques. By applying the following equation, we can determine the average performance $a_{(x,j)}$ of an algorithm denoted by $j=(1,2,3,\dots,m)$ in terms of the compression ratio, encoding time, and decoding time:
\begin{equation} \label{equ:eq4_6}
a_{(x,j)} = \frac{1}{n} \sum_{i=1}^N x_i \quad \text{for} \;  x_i \in \{r, e, d\}.
\end{equation}
To assess the effectiveness of a specific method relative to others, all average performances need to be mapped to a standard range. Additionally, the proportionality relationships \eqref{equ:eq1_3} must be taken into account. The resulting performance $c_{(x,j)}$ of algorithm $j$ according to performance measure $x$ can be formulated as follows: 

\begin{equation}\label{eq:eq7_9}
\begin{split}
c_{(x,j)} &\propto \frac{a_{(r,j)}}{\sum_{t=1}^m a_{(r,t)}} \text{ for } x = r; \\
c_{(x,j)} &\propto \frac{1}{a_{(x,j)}} \text{ for } x \in \{e,d\}.
\end{split}
\end{equation}

These relationships can be rewritten with constants of proportionality $k_r$, $k_e$, and $k_d$:
\begin{equation}\label{eq:eq10_12}
\begin{split}
c_{(x,j)} = k_x \times \left( \frac{\sum_{t=1}^m a_{(x,t)}}{a_{(x,j)}} \right) \text{ for } x = r; \\ \quad c_{(x,j)} = k_x \times \frac{1}{a_{(x,j)}} \text{ for } x\in \{e,d\}.
\end{split}
\end{equation}

Determining the optimal values for $k_r$, $k_e$, and $k_d$ is the most complex part of the procedure and is significant for the following reasons:
\begin{enumerate}
    \item Calculating the optimal values for these constants ensures a suitable balance between compression ratio, encoding time, and decoding time.    
    \item A good balance between these parameters assigns appropriate importance to each, providing the most effective strategy for a given situation.
    \item If the parameters are imbalanced, undue emphasis may be placed on one of the factors, leading to erroneous predictions.
\end{enumerate}
Let us choose $k_r$=1, $k_e = \sum_{v=1}^{m} a_{e,v}$, and $k_d = \sum_{v=1}^{m} a_{d,j}$. In this scenario, every compression method can have an impact on each performance metric. To prevent the results from being affected by a single compression method,  all values of  $c_{r,j}$, $c_{e,j}$, and $\ c_{d,j}$ must be averaged across all $m$ techniques. The following equation is used to eliminate any potential biases introduced by specific methods:
%
\begin{equation}\label{eq:eq13_15}
w = \frac{1}{m} \sum_{j=1}^{m} c_{x,j} \quad \text{for } x \in \{r, e, d\}
\end{equation}
where $w = x$ when $x = r$, $w = y$ when $x = e$, and $w = z$ when $x = d$.
When $x$ is substantially smaller than $y$ and $z$, inaccurate predictions may occur. Thus, the value of $x$ must be fine-tuned to align more closely with $y$ and $z$, which can be achieved by determining the optimal value for $k_r$.

Let us consider two variables $u$ and $v$. To attain a balance between the performance metrics, we use the following equations, which make adjustments to align $x$ with $y$ and $z$, respectively:
\begin{equation}\label{eq:eq16}
    x \times  u = y
\end{equation}
\begin{equation}\label{eq:eq17}
    x \times  v = z.
\end{equation}
However, it is crucial to ensure that any change in $x$ must have an equivalent effect on both $y$ and $z$. This is achieved by averaging $u$ and $v$, which enhances the balance between $x$, $y$, and $z$. Thus, $k_r$ is determined by
\begin{equation}\label{eq:eq18}
    k_r = \frac{{u + v}}{2} = \frac{{y + z}}{{2 \times x}}.
\end{equation}
By substituting the values of $k_r$, $k_e$, and $k_d$, equation (\ref{eq:eq10_12}) can be written as
\begin{equation}\label{eq:eq19_21}
\begin{split}
c_{x,j} = \frac{(y + z) \times a_{r,j}}{2 \times x \times \sum_{j=1}^{m} a_{r,j}} \text{ for } x = r; \\ \; c_{x,j} = \frac{\sum_{j=1}^{m} a_{x,j}}{a_{x,j}} \text{ for } x \in \{e, d\}.
\end{split}
\end{equation}

The parameters $r$, $e$, and $d$ are now balanced. We can now determine the grand total performance $g_j^{x_1,x_2}$ of the $j$th technique. When considering two factors, this is given by
\begin{equation}\label{eq:eq_22_24}
\begin{split}
g_j^{x_1,x_2} = \frac{ \sum_{k \in \{x_1,x_2\}} c_{k,j} }{ \sum_{i=1}^{m} \left( \sum_{k \in \{x_1,x_2\}} c_{k,j} \right) } \times 100, \\ \quad \text{where } x_1, x_2 \in \{r,e,d\} \text{ and } x_1 \neq x_2.
\end{split}
\end{equation}
Here, $x_1, x_2 \in \{r, e, d\}$ are the selected pair of factors. When considering all three factors, the grand total performance is
\begin{equation}\label{eq:eq25}
g_j^{r,e,d} = \frac{ \sum_{k \in \{r,e,d\}} c_{k,j} }{ \sum_{i=1}^{m} \left( \sum_{k \in \{r, e, d\}} c_{k,j} \right) } \times 100.
\end{equation}
Algorithm~\ref{algo:smn1} can now be applied to determine the most effective method \texttt{ml[sm]} along with its associated performance value \texttt{gp} for a given scenario. Here, \texttt{ml} denotes a list of all available methods, and \texttt{sm} is the index of a selected method.

\begin{algorithm}[!h]
    \caption{Finding the Optimal Method}
    \label{algo:smn1}
    \begin{algorithmic}[1]
        \Require \( g \), \( m \); Here, \( g \) represents the performance values of each method, and \( m \) is the total number of methods.
        \Ensure \texttt{SelectedMethod}, \texttt{GrandTotalPerformance}; Here, \texttt{SelectedMethod} is the method with the highest performance, \texttt{GrandTotalPerformance} is its associated performance value, \texttt{ml} denotes a list of all available methods, and \texttt{sm} is the index of the selected method.
        \State \( \texttt{gp} \gets -\infty \); Initialize \( \texttt{gp} \), the variable that stores the best performance value, to a very low number.
        \For{$i \gets 1$ to $m$}
            \State \( p \gets g_i \); Assign the performance value of the \(i\)th method to the variable \( p \).
            \If{$p > \texttt{gp}$}
                \State \( \texttt{gp} \gets p \); If the current performance \( p \) is greater than the previously stored best performance \( \texttt{gp} \), update \( \texttt{gp} \).
                \State \( \texttt{sm} \gets i \); Store the index \( i \) of the method that achieved the best performance in \( \texttt{sm} \).
            \EndIf
        \EndFor
        \State \Return \( \texttt{ml[sm]} \), \( \texttt{gp} \); Return the method corresponding to the index \( \texttt{sm} \) from the list of methods \( \texttt{ml} \), along with its best performance value \( \texttt{gp} \).
    \end{algorithmic}
\end{algorithm}

\section{Results and Analysis}\label{ResultsandAnalysis}
\begin{figure*}[!h]
    \centering
    \includegraphics[scale = .8]{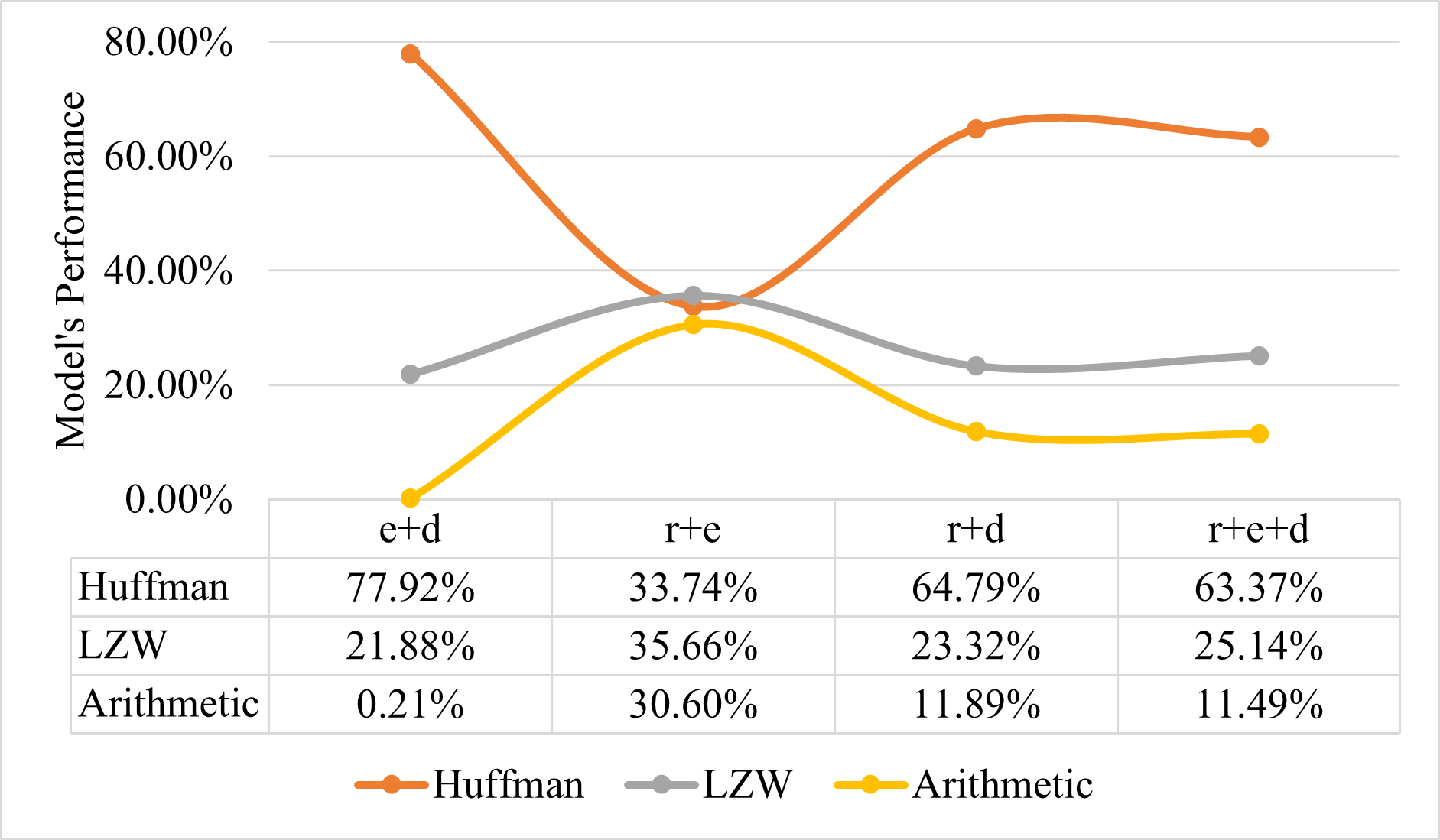}
    \caption{Outcome of the proposed approach for the results in Table~\ref{tab:my-table2}}
    \label{fig:Figure1}   
\end{figure*}

This section uses the proposed method to analyze recent findings in the literature. Performance measures for entropy coding techniques published in 2019 \cite{rahman2019lossless} are presented in Table~\ref{tab:my-table2}. These show that the Huffman and LZW techniques perform encoding and decoding faster than Arithmetic. However, compared with Huffman, LZW achieves a slightly higher compression ratio, making it a better choice for applications in which storage space is a critical factor. The results of the proposed technique in this study are evaluated using the data in Table~\ref{tab:my-table2}. Fig.~\ref{fig:Figure1}  illustrates the outcomes.

\begin{table}[!h]
\centering
\caption{Outcomes of entropy coding techniques \cite{rahman2019lossless}}
\label{tab:my-table2}
\begin{tabular}{|c|c|c|c|}
\hline
\textbf{Compressor} &
  \textbf{\begin{tabular}[c]{@{}c@{}}($e$)\end{tabular}} &
  \textbf{\begin{tabular}[c]{@{}c@{}}($d$)\end{tabular}} &
  \textbf{\begin{tabular}[c]{@{}c@{}}($r$)\end{tabular}} \\ \hline
Huffman    & 0.2488 & 0.0062 & 1.4282 \\ \hline
LZW        & 0.1054 & 0.0266 & 1.5455 \\ \hline
Arithmetic & 4.0178 & 4.7208 & 1.4380 \\ \hline
\end{tabular}
\end{table}

Fig.~\ref{fig:Figure1} shows a comparison of three entropy coding techniques, Huffman, LZW, and Arithmetic, using different combinations of parameters. When considering encoding time and decoding time, as well as encoding time and compression ratio, Huffman and LZW achieve the best results, with 77.92\% and 35.66\%, respectively. Conversely, when combining decoding time and compression ratio or when considering all three parameters, Huffman performs best, with 64.79\% and 63.37\%, respectively. Therefore, LZW is a superior choice if the encoding speed and compression ratio are the primary concerns. However, if other factors are also considered, Huffman coding is the preferred entropy coding technique.

\begin{figure*}[!h]
    \centering
    \includegraphics[scale = .80]{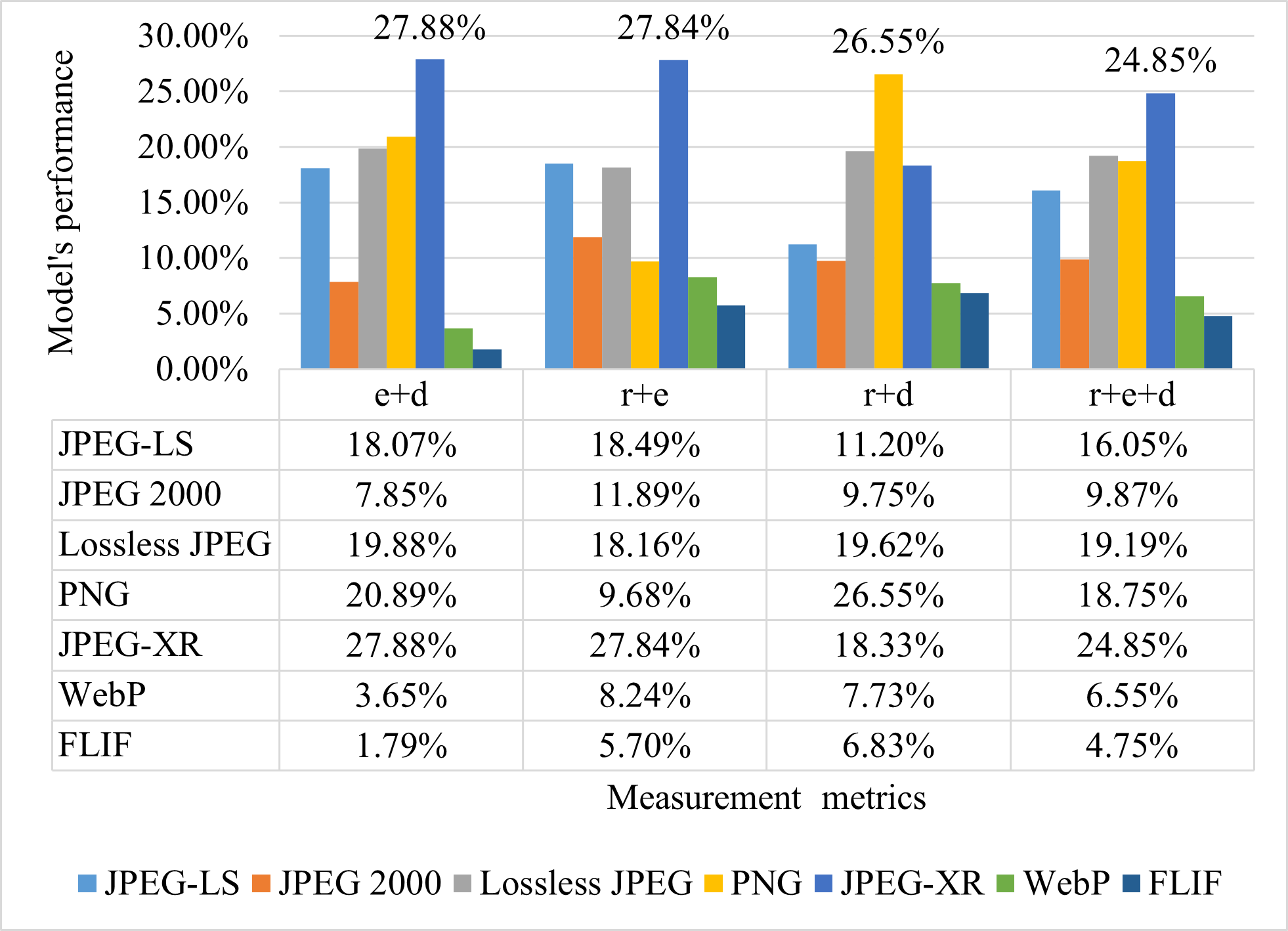}
    \caption{Outcome of the proposed approach for the results in Table~\ref{tab:my-table3}}
    \label{fig:Figure2}   
\end{figure*}

Table~\ref{tab:my-table3} presents the results of research published in 2021 \cite{rahman2021impact}. In this case, FLIF achieves the highest compression ratio among all techniques, with a value of 4.4508. JPEG-XR and PNG exhibit the fastest encoding and decoding times, respectively. Specifically, PNG has an encoding time of 1.1259 seconds and a decoding time of 0.0719 seconds, whereas JPEG-XR has an encoding time of 0.1668 seconds and a decoding time of 0.1614 seconds. The results of the proposed approach for the data in Table~\ref{tab:my-table3} are shown in Fig.~\ref{fig:Figure2}.

\begin{table}[!h]
\centering
\caption{Outcomes of transform techniques \cite{rahman2021impact}}
\label{tab:my-table3}
\begin{tabular}{|l|c|c|c|}
\hline
\multicolumn{1}{|c|}{\textbf{Compressor(C)}} &
  \textbf{\begin{tabular}[c]{@{}c@{}}($e$)\end{tabular}} &
  \textbf{\begin{tabular}[c]{@{}c@{}}($d$)\end{tabular}} &
  \textbf{\begin{tabular}[c]{@{}c@{}}($r$)\end{tabular}} \\ \hline
JPEG-LS       & 0.2467 & 0.279  & 3.9616 \\ \hline
JPEG 2000     & 0.597  & 0.563  & 3.4189 \\ \hline
Lossless JPEG & 0.3257 & 0.1331 & 2.6661 \\ \hline
PNG           & 1.1259 & 0.0719 & 3.2847 \\ \hline
JPEG-XR       & 0.1668 & 0.1614 & 2.5307 \\ \hline
WebP          & 1.33   & 1.1156 & 3.8517 \\ \hline
FLIF          & 5.1033 & 1.1354 & 4.4508 \\ \hline
\end{tabular}
\end{table}

Fig.~\ref{fig:Figure2} displays percentages for various combinations of encoding time, decoding time, and compression ratio for several image compression techniques, namely, JPEG-LS, JPEG 2000, lossless JPEG, PNG, JPEG-XR, WebP, and FLIF. The percentages reflect the relative importance of these factors when combined. When considering decoding time and compression ratio, PNG is the optimal choice, but JPEG-XR is the best option for all other combinations.

\begin{figure*}[!h]
    \centering
    \includegraphics[scale = .85]{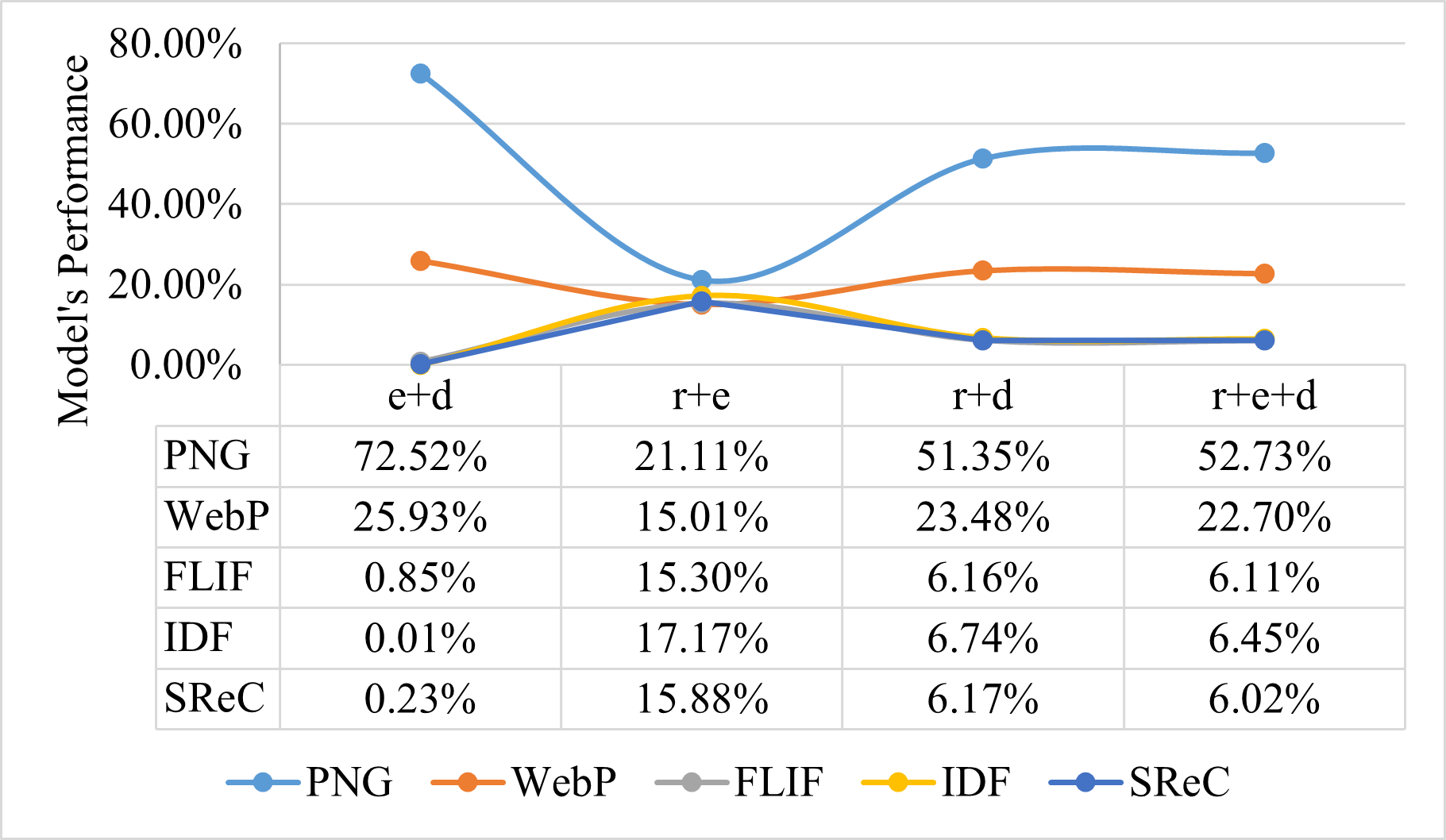}
    \caption{Outcome of the proposed approach for the results in Table~\ref{tab:my-table4}}
    \label{fig:Figure3}   
\end{figure*}

Bits per subpixel (bpsp) measures the number of bits required to represent each color component of a single pixel in an image. Each pixel is divided into three smaller subpixels—red, green, and blue—each of which can be represented by 8 bits. Table~\ref{tab:my-table4}  summarizes the results of a study by Cao et al. \cite{cao2020lossless} on the ImageNet64 \cite{chrabaszcz2017downsampled} dataset, comparing several classical and learning-based methods. The bpsp values are converted to compression ratios ($r$) using
\begin{equation}\label{eq:eq26}
    r = \frac{8}{\text{bpsp}}.
\end{equation}
 Although PNG encodes and decodes data faster than the other methods, IDF achieves a higher $r$ value for this dataset. To identify the most suitable compression method for a particular application, the proposed approach can be applied to the results in Table~\ref{tab:my-table4}, with the results shown in Fig.~\ref{fig:Figure3}. It can be seen that PNG is the most effective for all combinations. Therefore, for this dataset, PNG appears to be the most suitable choice unconditionally.

\begin{table}[!h]
\centering
\caption{Outcomes of some learning and non-learning compressors \cite{cao2020lossless}}
\label{tab:my-table4}
\begin{tabular}{|c|c|c|c|}
\hline
\textbf{C} &
  \textbf{\begin{tabular}[c]{@{}c@{}} ($e$)\end{tabular}} &
  \textbf{\begin{tabular}[c]{@{}c@{}}($d$)\end{tabular}} &
  \textbf{\begin{tabular}[c]{@{}c@{}}($r$)\end{tabular}} \\ \hline
PNG  & 0.0013 & 0.00008 & 1.3937 \\ \hline
WebP & 0.021  & 0.00021 & 1.7241 \\ \hline
FLIF & 0.022  & 0.01    & 1.7621 \\ \hline
L3C  & 0.031  & 0.023   & 1.81   \\ \hline
IDF  & 1.33   & 1.02    & 2.0513 \\ \hline
SReC & 0.044  & 0.071   & 1.8648 \\ \hline
\end{tabular}
\end{table}

\begin{figure*}[!h]
    \centering
    \includegraphics[scale = .6]{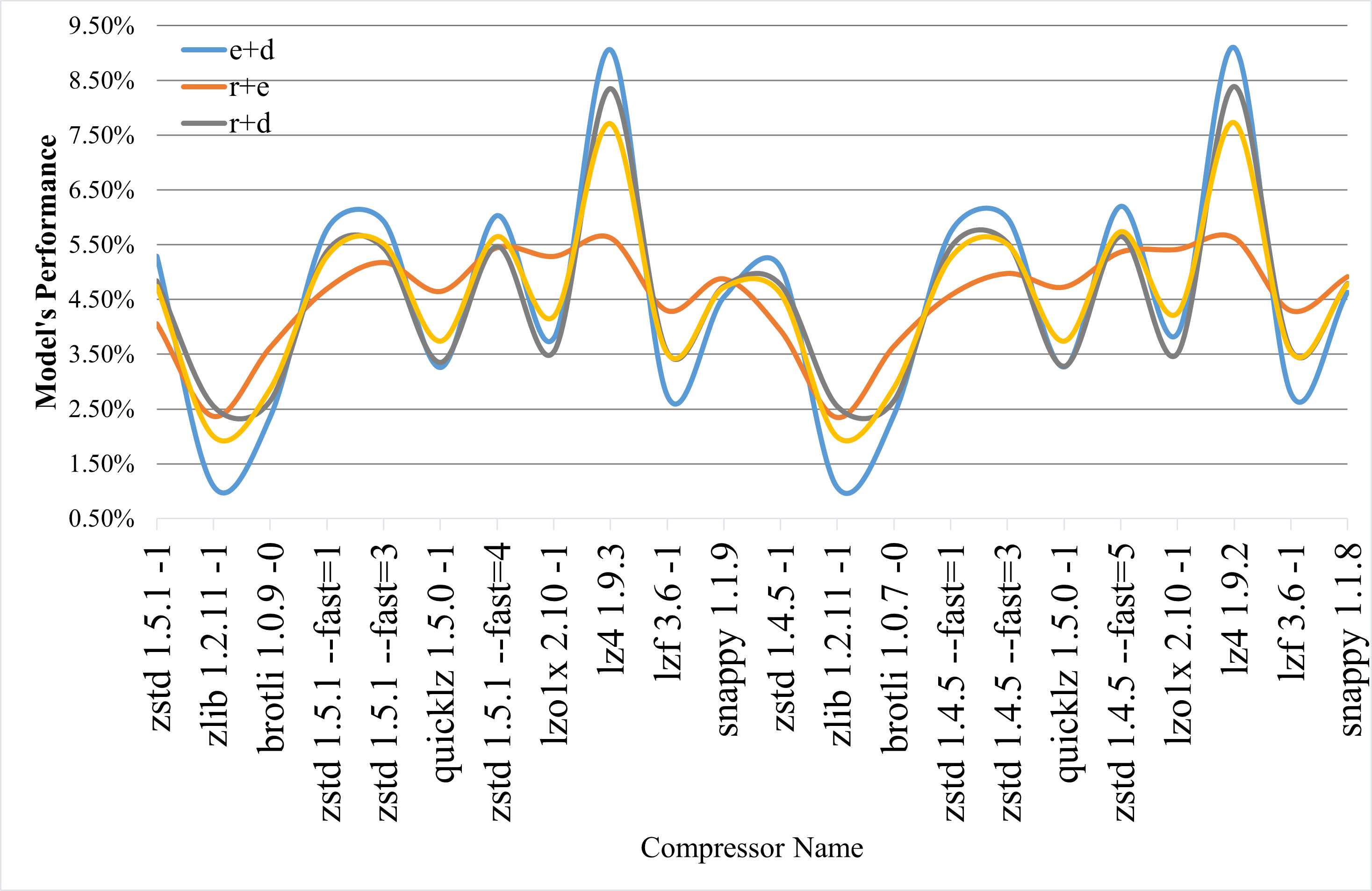}
    \caption{Outcome of the proposed technique for the results in Table~\ref{tab:my-table5}}
    \label{fig:Figure4}   
\end{figure*}

The Silesia compression corpus \cite{REFERENCELINK7} contains 211,938,580 bytes. Researchers at Facebook \cite{TransactionLink5, TransactionLink6} applied several text compressors to this corpus and calculated the compression ratios, as well as the compression speed (CS) and decompression speed (DS)  in megabytes per second. The following equations are used to convert the compression and decompression speeds into the encoding time $e$ and decoding time $d$, given the compression ratio $r$:
\begin{equation}\label{eq:eq27}
    e = \frac{\text{size [bytes]}}{\text{CS [MB/s]} \times 1024 \times 1024}
\end{equation}
\begin{equation}\label{eq:eq28}
    d = \frac{\text{size [bytes]}}{\text{DS[MB/s]} \times r \times 1024 \times 1024}.
\end{equation}
The results are presented in Table~\ref{tab:my-table5}. Fig.~\ref{fig:Figure4} illustrates the outcomes of applying the proposed approach to the data in Table~\ref{tab:my-table5}.

\begin{table}[!h]
\centering
\caption{Outcomes of text compressors on the Silesia compression corpus \cite{REFERENCELINK7}}
\label{tab:my-table5}
\begin{tabular}{|l|c|c|c|}
\hline
\textbf{Compressor} &
  \textbf{\begin{tabular}[c]{@{}c@{}}($e$)\end{tabular}} &
  \textbf{\begin{tabular}[c]{@{}c@{}}($d$)\end{tabular}} &
  \textbf{\begin{tabular}[c]{@{}c@{}}($r$)\end{tabular}} \\ \hline
zstd 1.5.1 \texttt{-1}       & 0.0215 & 0.0023 & 2.887 \\ \hline
zlib 1.2.11 \texttt{-1}      & 0.1198 & 0.0104 & 2.743 \\ \hline
brotli 1.0.9 \texttt{-0}     & 0.0288 & 0.0094 & 2.702 \\ \hline
zstd 1.5.1 \texttt{--fast=1} & 0.019  & 0.0022 & 2.437 \\ \hline
zstd 1.5.1 \texttt{--fast=3} & 0.017  & 0.0023 & 2.239 \\ \hline
quicklz 1.5.0 \texttt{-1}    & 0.0211 & 0.0067 & 2.238 \\ \hline
zstd 1.5.1 \texttt{--fast=4} & 0.016  & 0.0023 & 2.148 \\ \hline
lzo1x 2.10 \texttt{-1}       & 0.0173 & 0.0064 & 2.106 \\ \hline
lz4 1.9.3           & 0.0154 & 0.0012 & 2.101 \\ \hline
lzf 3.6 \texttt{-1}          & 0.0278 & 0.0066 & 2.077 \\ \hline
snappy 1.1.9        & 0.0207 & 0.0031 & 2.073 \\ \hline
zstd 1.4.5 \texttt{-1}       & 0.0228 & 0.0024 & 2.884 \\ \hline
zlib 1.2.11 \texttt{-1}      & 0.1265 & 0.0104 & 2.743 \\ \hline
brotli 1.0.7 \texttt{-0}     & 0.0285 & 0.0094 & 2.703 \\ \hline
zstd 1.4.5 \texttt{--fast=1} & 0.02   & 0.0021 & 2.434 \\ \hline
zstd 1.4.5 \texttt{--fast=3} & 0.0178 & 0.0021 & 2.312 \\ \hline
quicklz 1.5.0 \texttt{-1}    & 0.0203 & 0.0072 & 2.238 \\ \hline
zstd 1.4.5 \texttt{--fast=5} & 0.0163 & 0.0022 & 2.178 \\ \hline
lzo1x 2.10 \texttt{-1}       & 0.0165 & 0.0066 & 2.106 \\ \hline
lz4 1.9.2           & 0.0154 & 0.0012 & 2.101 \\ \hline
lzf 3.6 \texttt{-1}          & 0.0278 & 0.0064 & 2.077 \\ \hline
snappy 1.1.8        & 0.0203 & 0.0031 & 2.073 \\ \hline
\end{tabular}
\end{table}

From the results presented in Fig.~\ref{fig:Figure4}, it can be seen that LZ4 1.9.2 outperforms all other methods with percentages of 9.10\%, 5.63\%, 8.39\%, and 7.73\% for combinations of $e$ and $d$, $e$ and $r$, $d$ and $r$, and all three parameters, respectively. Therefore, LZ4 1.9.2 is the optimal method for this dataset, regardless of how the compression criteria are combined.

\section{Validation}\label{Validatetheproposedmethod}
The purpose of this section is to validate the proposed method with evidence. To achieve this, five hypothetical text compressors (C), designated as C1 through C5, are considered. Tables~\ref{tab:my-table6}–\ref{tab:my-table9} display the results with random values assigned to the encoding time $e$, decoding time $d$, and compression ratio $r$. Specifically, for the combinations $e$ and $d$, $e$ and $r$, $d$ and $r$, and all three parameters, we deliberately select lower values for compressors C2, C3, C5, and C4, respectively, to evaluate the effectiveness of the proposed method.

\begin{table}[!h]
\centering
\caption{X1 Dataset}
\label{tab:my-table6}
\begin{tabular}{|c|c|c|c|}
\hline
\textbf{C} &
  \textbf{\begin{tabular}[c]{@{}c@{}}($e$)\end{tabular}} &
  \textbf{\begin{tabular}[c]{@{}c@{}}($d$)\end{tabular}} &
  \textbf{\begin{tabular}[c]{@{}c@{}}($r$)\end{tabular}} \\ \hline
C1 & 0.0727          & 0.0233          & 1.885 \\ \hline
C2 & \textbf{0.0722} & \textbf{0.0227} & 4.739 \\ \hline
C3 & 0.0751          & 0.0252          & 7.992 \\ \hline
C4 & 0.1985          & 0.0603          & 9.334 \\ \hline
C5 & 0.0823          & 0.0294          & 3.414 \\ \hline
\end{tabular}
\end{table}

\begin{table}[!h]
\caption{X2 Dataset}
\label{tab:my-table7}
\begin{tabular}{|c|c|c|c|}
\hline
\textbf{C} &
  \begin{tabular}[c]{@{}c@{}}($e$)\end{tabular} &
  \begin{tabular}[c]{@{}c@{}}$d$\end{tabular} &
  \begin{tabular}[c]{@{}c@{}}$r$\end{tabular} \\ \hline
C1 & 0.0727          & 0.0233 & 1.885          \\ \hline
C2 & 0.0722          & 0.0227 & 4.739          \\ \hline
C3 & \textbf{0.0711} & 0.0252 & \textbf{4.792} \\ \hline
C4 & 0.1985          & 0.0603 & 4.334          \\ \hline
C5 & 0.0823          & 0.0294 & 3.414          \\ \hline
\end{tabular}
\end{table}

\begin{table}[!h]
\caption{X3 Dataset}
\label{tab:my-table8}
\begin{tabular}{|c|c|c|c|}
\hline
\textbf{C} &
  \textbf{\begin{tabular}[c]{@{}c@{}}($e$)\end{tabular}} &
  \textbf{\begin{tabular}[c]{@{}c@{}}($d$)\end{tabular}} &
  \textbf{\begin{tabular}[c]{@{}c@{}}($r$)\end{tabular}} \\ \hline
C1 & 0.0727 & 0.0233          & 1.885          \\ \hline
C2 & 0.0722 & 0.0227          & 4.739          \\ \hline
C3 & 0.0711 & 0.0252          & 4.792          \\ \hline
C4 & 0.1985 & 0.0603          & 4.334          \\ \hline
C5 & 0.0823 & \textbf{0.0214} & \textbf{4.814} \\ \hline
\end{tabular}
\end{table}

\begin{table}[!h]
\caption{X4 Dataset}
\label{tab:my-table9}
\begin{tabular}{|c|c|c|c|}
\hline
\textbf{C} &
  \textbf{\begin{tabular}[c]{@{}c@{}}($e$)\end{tabular}} &
  \textbf{\begin{tabular}[c]{@{}c@{}}($d$)\end{tabular}} &
  \textbf{\begin{tabular}[c]{@{}c@{}}($r$)\end{tabular}} \\ \hline
C1 & 0.0727          & 0.0233          & 1.885          \\ \hline
C2 & 0.0722          & 0.0227          & 4.739          \\ \hline
C3 & 0.0711          & 0.0252          & 4.792          \\ \hline
C4 & \textbf{0.0708} & \textbf{0.0203} & \textbf{4.834} \\ \hline
C5 & 0.0823          & 0.0214          & 4.814          \\ \hline
\end{tabular}
\end{table}

\begin{figure}[!h]
    \centering
    \includegraphics[width=0.8\textwidth]{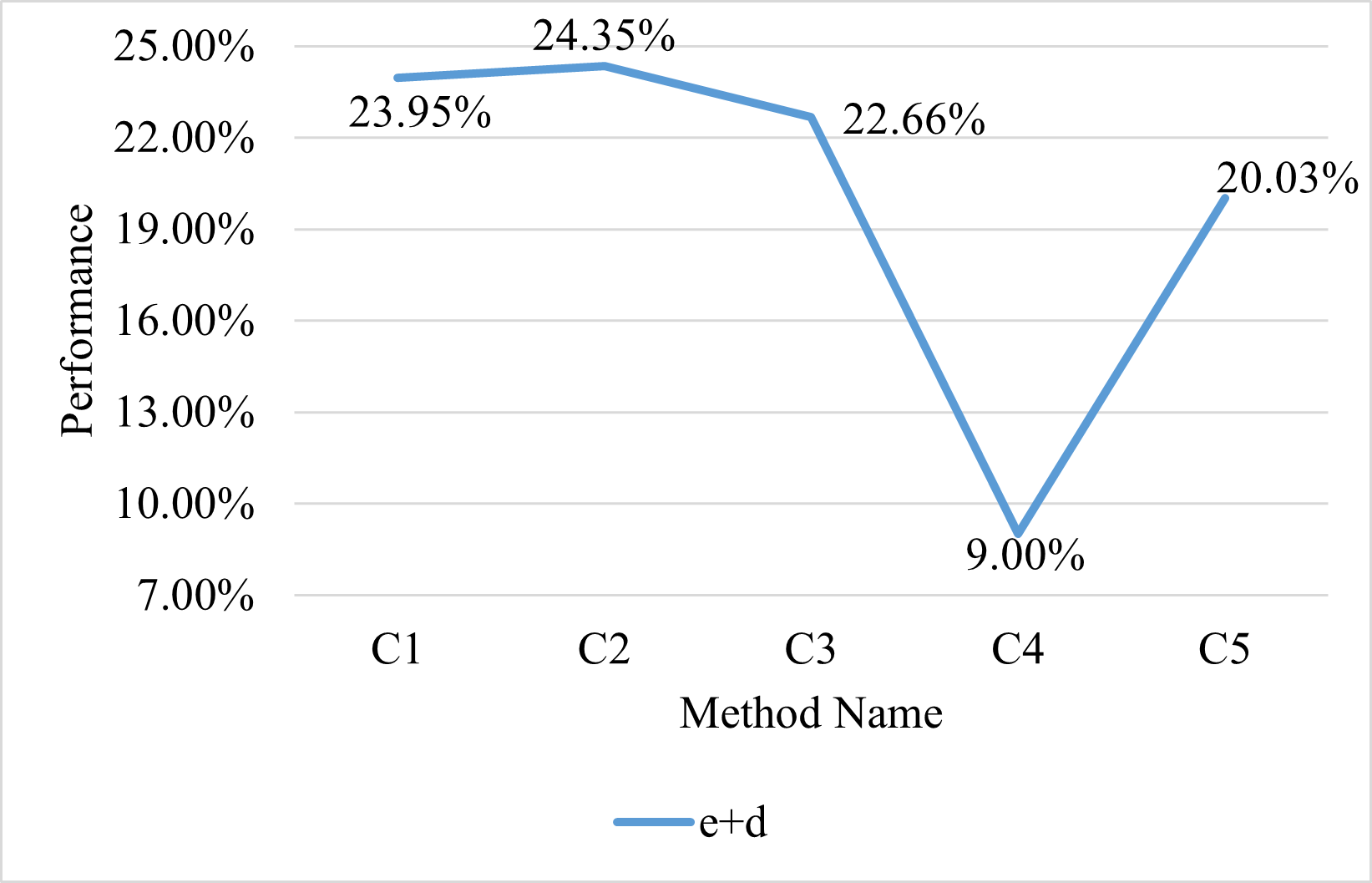}
    \caption{\footnotesize Performance for the combination of $e$ and $d$.}
    \label{fig:Figure5}
\end{figure}

\begin{figure}[!h]
    \centering
    \includegraphics[width=0.8\textwidth]{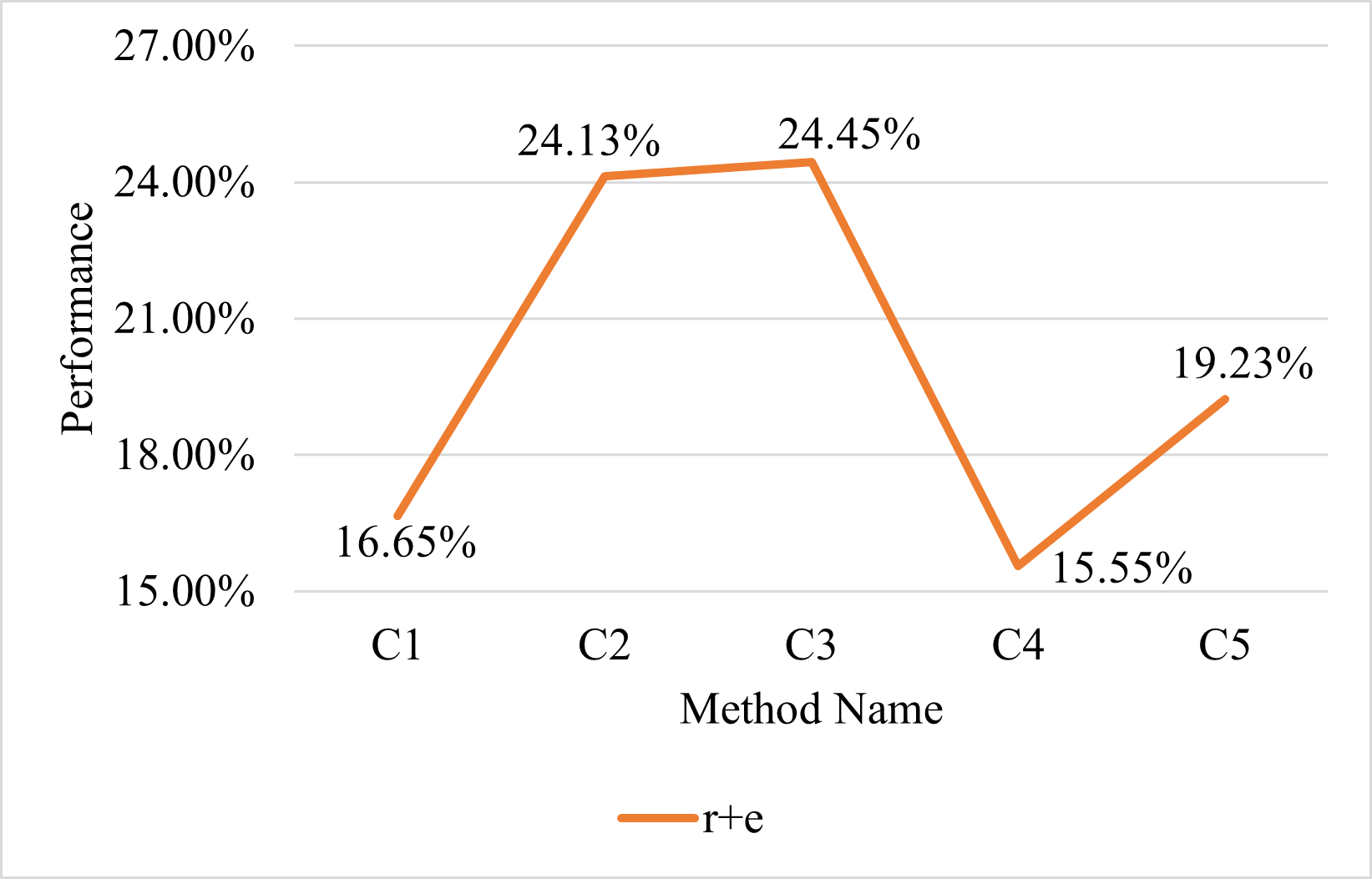}
    \caption{\footnotesize Performance for the combination of $r$ and $e$.}
    \label{fig:Figure6}
\end{figure}

\begin{figure}[!h]
    \centering
    \includegraphics[width=0.8\textwidth]{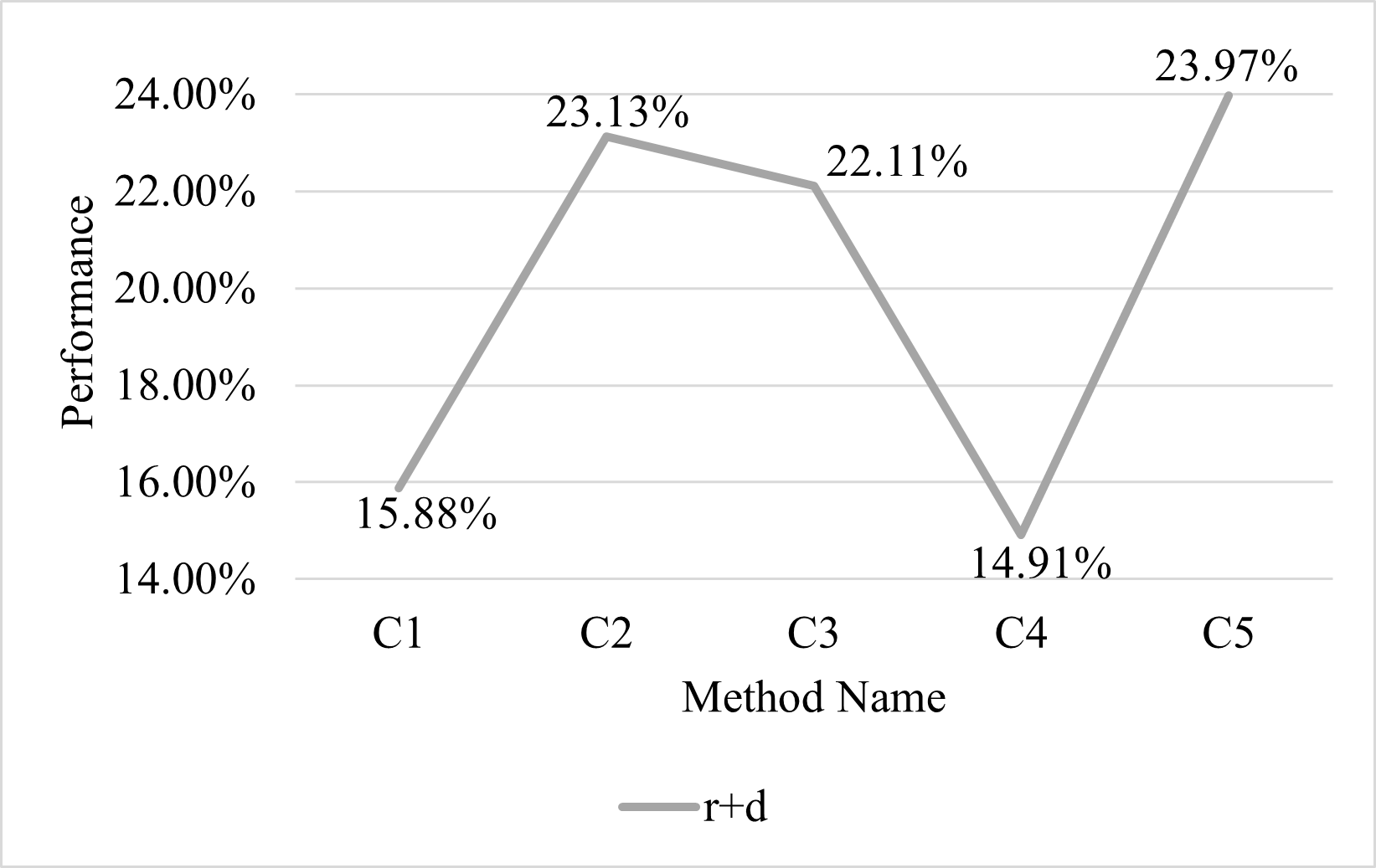}
    \caption{\footnotesize Performance for the combination of $d$ and $r$.}
    \label{fig:Figure7}
\end{figure}

\begin{figure}[!h]
    \centering
    \includegraphics[width=0.8\textwidth]{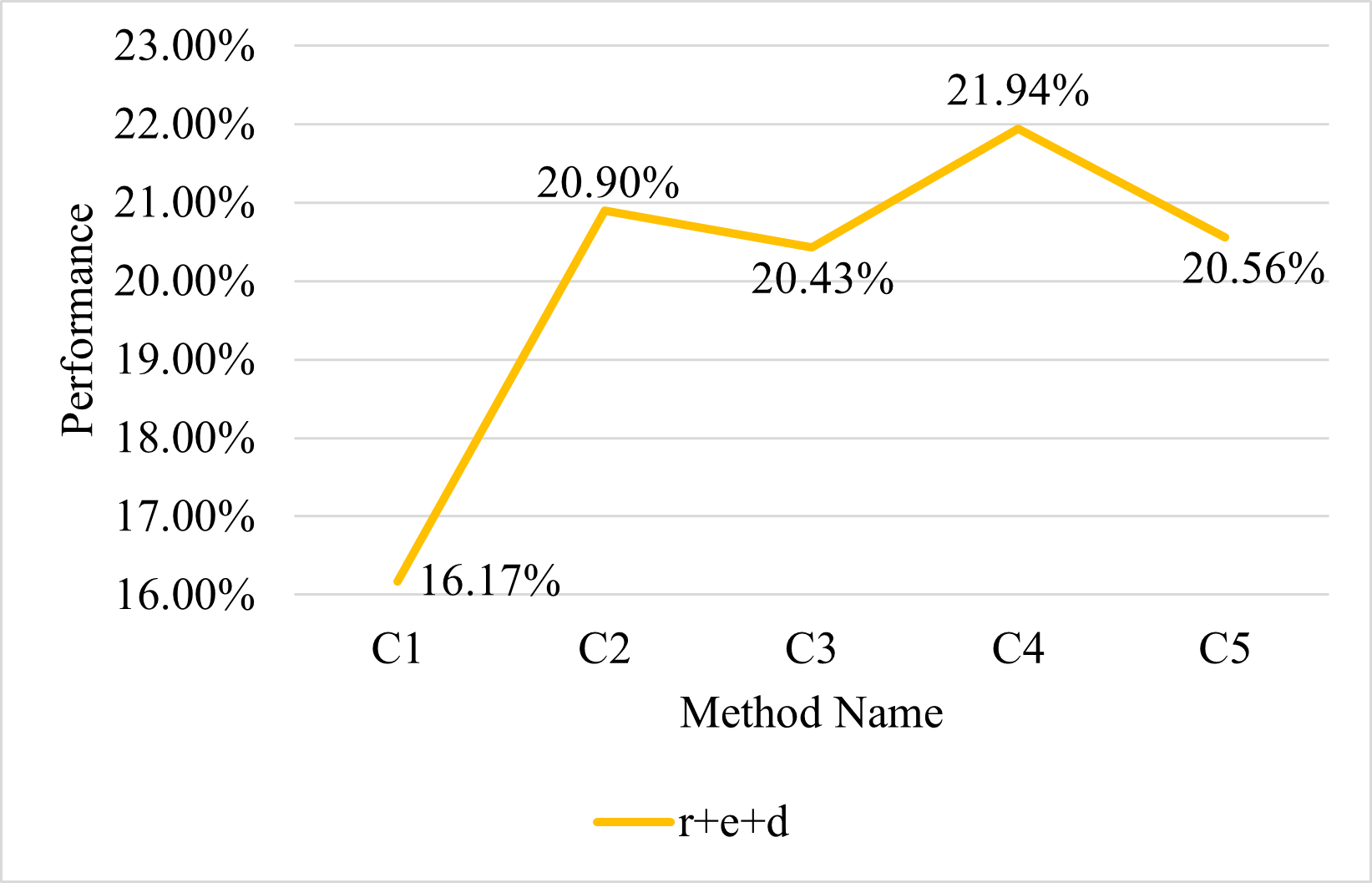}
    \caption{\footnotesize Performance for the combination of $e$, $d$, and $r$.}
    \label{fig:Figure8}
\end{figure}

The results of our proposed strategy in Tables~\ref{tab:my-table6}–\ref{tab:my-table9} are illustrated in Figs.~\ref{fig:Figure5}–\ref{fig:Figure8}, which demonstrate that the most appropriate approaches are correctly identified. Therefore, we can conclude that our method effectively predicts the optimal compressor for any combination of $e$, $d$, and $r$.

\section{Conclusion}\label{conclusions1}
The objective of this article is to introduce a method for identifying the most efficient lossless compression algorithm in a particular application. The effectiveness of a lossless compression algorithm relies on all parameters—compression ratio, encoding time, and decoding time—rather than any single parameter alone. Although prior research has predominantly evaluated algorithms on the basis of individual parameters, the performance in a given application typically depends on two or all of them. Consequently, selecting an optimal lossless data compression algorithm for a given set of application requirements is a challenging task.

To address this challenge, we propose a model that predicts the optimal lossless data compression algorithm for any combination of these parameters and quantifies the overall performance of each compared algorithm as a percentage. Experimental findings confirm that the proposed model effectively predicts the optimal method across various parameter combinations. Furthermore, the results highlight that although learning-based methods offer high compression ratios, classical methods generally exhibit greater efficacy for lossless compression.

\bibliography{sn-bibliography}
\end{document}